# Highly efficient spin current generation by the spin Hall effect in Au$_{1-x}$Pt$_x$

Lijun Zhu,[1] Daniel. C. Ralph, [1,2] Robert A. Buhrman[1]
1. Cornell University, Ithaca, NY 14850
2. Kavli Institute at Cornell, Ithaca, New York 14853, USA

We report very efficient spin current generation by the spin Hall effect in the alloy Au$_{0.25}$Pt$_{0.75}$, which, as determined by two different direct spin-orbit torque measurements, exhibits a giant internal spin Hall ratio of ≥ 0.58 (anti-damping spin-orbit torque efficiency of ≈ 0.35 in bilayers with Co), a relatively low resistivity of ≈ 83 μΩ cm, an exceptionally large spin Hall conductivity of ≥ 7.0×10$^5$ Ω$^{-1}$ m$^{-1}$, and a spin diffusion length of 1.7 nm. This work establishes Au$_{0.25}$Pt$_{0.75}$ as a milestone spin current generator that provides greater energy efficiency than that yet obtained with other heavy metals or with the topological insulators Bi$_2$Se$_3$ and (Bi,Se)$_2$Te$_3$. Our findings should advance spin-orbit torque-based fundamental research and benefit the development of new fast, efficient spin-orbit torque-driven magnetic memories, skyrmion and chiral domain wall devices, and microwave and terahertz emitters.

## I. INTRODUCTION

Current-induced spin-orbit torques (SOTs) in heavy metal/ferromagnet (HM/FM) systems have attracted considerable attention due to their potential for application in the efficient manipulation of magnetization at the nanoscale [1-6]. The pure transverse spin current density $J_s$ arising from an applied longitudinal electrical density $J_e$ via the spin Hall effect (SHE) in HMs can generate damping-like SOT sufficiently strong to excite magnetization dynamics at microwave or terahertz frequencies [3,4], create and move skyrmions [5], drive chiral domain wall displacement [6], or switch the magnetization of thin film nanomagnets [1,2]. Of particular technological promise are the HMs with a large spin Hall ratio ($\theta_{SH}$), or more precisely with a large damping-like SOT efficiency $\xi_{DL} \equiv T_{int}\theta_{SH} \equiv T_{int}(2e/\hbar)J_s/J_e$, which is exerted on an adjacent ferromagnetic layer, as these could allow for the development of very fast, deterministic and efficient magnetic random access memories (MRAMs). Here $T_{int}$ (<1) is the spin transparency of the HM/FM interface. For example, the SHE in Pt (W) has been demonstrated recently to switch in-plane magnetic tunnel junctions with a critical switching current density of 4.0 (0.54) ×10$^7$ A/cm$^2$ and a write error rate below 10$^{-5}$ (10$^{-6}$) with a 2 ns pulse duration [7,8]. Despite extensive efforts, the efficiency of present SOT operations are still limited by a relatively low $\xi_{DL}$ and/or a very high resistivity ($\rho_{xx}$) of the HM. For instance, β-W has a large |$\xi_{DL}$| of ~0.3 but is very resistive ($\rho_{xx}$ ≈ 300 μΩ cm at 4 nm)[9]; |$\xi_{DL}$| generally reported for Pt and β-Ta is comparatively low (~ -0.12), and β-Ta is also very resistive (≈180 μΩ cm)[1,10]. These factors lead to undesirably large energy dissipation.

It has been established that the SHE in Pt is generally dominated by the intrinsic band structure effect [11,12] with the spin Hall conductivity $\sigma_{SH} \equiv \sigma_{SH}^{intr} + \sigma_{SH}^{extr} \approx \sigma_{SH}^{intr}$. Here $\sigma_{SH}^{intr}$ and $\sigma_{SH}^{extr}$ are the intrinsic and extrinsic contributions. Hence $\xi_{DL}$ (= $T_{int}(2e/\hbar)\sigma_{SH}\rho_{xx}$) can be enhanced by raising $\rho_{xx}$ provided that $\sigma_{SH}$ is not degraded in the process. Some success has been achieved in this manner by alloying Pt with Hf or with Al, but the maximum $\xi_{DL}$ is still below 0.16 even when $\rho_{xx}$ is enhanced to 110 μΩ cm, presumably due to a decrease in $\sigma_{SH}$ with increasing Hf or Al concentration [10]. Recently theoretical work has suggested that alloying Pt with Au would be particularly effective in increasing $\xi_{DL}$ via raising $\rho_{xx}$, due to a slower reduction in $\sigma_{SH}$ at high Pt concentration [13].

Here we report from direct SOT measurements that the Au$_{1-x}$Pt$_x$ alloy is indeed an enhanced spin current generator, with a maximum $\xi_{DL}$ of 0.35 while still maintaining a relatively low $\rho_{xx}$ of ≈ 83 μΩ cm and a relatively long spin diffusion length ($\lambda_s$) of 1.7 nm for the optimized composition $x$ = 0.75. We also determined a lower bound for $\theta_{SH}$ of 0.58 and a lower bound for the $\sigma_{SH}$ of 7.0×10$^5$ Ω$^{-1}$ m$^{-1}$ for Au$_{0.25}$Pt$_{0.75}$. While the variation of $\xi_{DL}$ with Pt concentration $x$ is qualitatively similar in form to the theoretical prediction both $\theta_{SH}$ and $\sigma_{SH}$ are much greater, approximately twice, than both the theoretical prediction and the experimental results from the indirect spin pumping/inverse SHE measurements reported in Ref. [13].

## II. RESULTS AND DISCUSSIONS
### A. Samples and characterizations

The samples for this work include two series of multilayer stacks: (I) Ta 1.0/Au$_{1-x}$Pt$_x$ 4.0 /Co 0.8-1.4/MgO 2.0/Ta 1.5 (numbers are thickness in nm) with $x$ = 0, 0.25, 0.5, 0.65, 0.75, 0.85, and 1, respectively; (II) Ta 1.0/Au$_{0.25}$Pt$_{0.75}$ 0, 2, 3, 4, 5, 6, and 8/Co 0.8-1.4/MgO 2.0/Ta 1.5 (see Fig. 1(a)). All the stacks were sputtered at room temperature on SiO$_2$/Si wafers with argon pressure of 2 mTorr and a base pressure of below 1×10$^{-8}$ Torr. The 1 nm Ta underlayer was introduced prior to co-sputtering of to improve the adhesion and uniformity of the Au$_{1-x}$Pt$_x$ layer but is expected to contribute negligible spin current into the Co layer as a consequence of the small thickness and high $\rho_{xx}$ of Ta, the short $\lambda_s$ of Au$_{1-x}$Pt$_x$, and the opposite signs of $\xi_{DL}$ for Ta and Au$_{1-x}$Pt$_x$. The 1.5 nm Ta capping layer was fully oxidized upon exposure to atmosphere. X-ray diffraction studies show that, as is the case with the elemental metal Au and Pt, Au$_{1-x}$Pt$_x$ has a face-centered-cubic structure with the Au$_{1-x}$Pt$_x$ (111) peak shifting from Pt to Au with increasing Au concentration (Fig. 1(b)), indicating the variation of lattice constant according to Vegard's Law (Fig. 1(c)). Vibrating sample magnetometer measurements as a function of Co thicknesses ($t$) show that there is no significant magnetic dead layer at the Au$_{1-x}$Pt$_x$/Co and Co/MgO interfaces and that the Co layers have an almost $x$-independent saturation magnetization ($M_s$) of 1413 ± 82 emu/cc [14], which is slightly smaller than the Co bulk value of ~1450 emu/cc and indicative of the absence of a significant magnetic proximity effect [15].

As shown in Fig. 1(d), the stacks were further patterned into 5×60 μm² Hall bars by ultraviolet photolithography and argon ion milling. The Au$_{1-x}$Pt$_x$ resistivity for each composition ($x$) and thickness ($d$) was determined by measuring the conductance enhancement of the corresponding stacks ($d > 0$ nm) with respect to the $d = 0$ nm stack. The Co resistivity was determined to be ~80 and 65 μΩ cm for 0.85 and 1.4 nm thicknesses, respectively, higher than bulk Co due to the strong interfacial scattering. For harmonic response measurements, a lock-in amplifier was used to source a sinusoidal voltage ($V_{in} = 4$ V) onto the 60 μm-long bar orientated along the $x$ axis and to detect the in-phase first and out-of-phase second harmonic Hall voltages, $V_{1\omega}$ and $V_{2\omega}$. The direct current switching of magnetization was measured with dc current sourced by a Yokogawa 7651 and with the differential Hall resistance detected by the lock-in amplifier ($V_{in} = 0.1$ V).

### B. Composition dependence of Spin-orbit torques

Figure 1(e) shows three examples of the anomalous Hall voltage hysteresis loops of Au$_{1-x}$Pt$_x$ 4 nm/Co 0.8 nm bilayers with $x = 1$, 0.75, and 0.25, where the fairly square loops and large $H_c$ provide evidence of the strong perpendicular magnetic anisotropy (PMA) of these samples. All the samples with $x \geq 25\%$ exhibit good PMA, $H_c$ larger than 500 Oe, and perpendicular anisotropy field $H_k$ of 6-12 [14]. The strong PMA allows us to obtain high-quality out-of-plane harmonic response data. Pure Au/Co bilayers do not show PMA in the studied Co thickness range. We carried out in-plane harmonic response measurement on Au$_{1-x}$Pt$_x$ 4 /Co 1.4 bilayers with in-plane magnetic anisotropy (IMA) to reaffirm the SOT efficiencies for each $x$ from 0 to 1.

We first discuss the SOTs in 4 nm Au$_{1-x}$Pt$_x$ samples with different $x$. Figure 2(a) shows an example of $V_{1\omega}$ and $V_{2\omega}$ to an applied ac current for perpendicularly magnetized Au$_{1-x}$Pt$_x$ 4/Co 0.8 bilayers ($x \geq 0.25$) as a function of in-plane bias fields $H_x$ and $H_y$, from which the damping-like and field-like effective spin-torque fields are determined by $H_{DL(FL)} = -2\frac{\partial V_{2\omega}}{\partial H_{x(y)}} / \frac{\partial^2 V_{1\omega}}{\partial^2 H_{x(y)}}$. As summarized in Fig. 2(c), both $H_{DL}$ and $H_{FL}$ are tuned significantly by $x$. Figure 2(d) shows $\xi_{DL}$ and field-like SOT efficiency $\xi_{FL}$ calculated by $\xi_{DL(FL)} = 2e\mu_0 M_s t H_{DL(FL)}/\hbar j_e$, with $e$, $\mu_0$, $t$, $\hbar$, and $j_e$ being the elementary charge, the permeability of vacuum, the ferromagnetic layer thickness, the reduced Planck constant, and the charge current density, respectively. Most notably, $\xi_{DL}$ climbs up quickly from 0.03 at $x = 0$ (pure Au) to the peak value of 0.30 at $x = 0.75$ and then gradually drops to 0.18 at $x = 1$ (pure Pt). We consistently observed the same $x$ dependence with in-plane harmonic measurements [14] of $\xi_{DL}$ on Au$_{1-x}$Pt$_x$ 4/Co 1.4 IMA samples (Fig. 2(d)). We note here that $\xi_{DL}$ of 0.18 for the 4 nm pure Pt sample is larger than 0.12 as previously reported for Pt/Co bilayers [16], but in that case $\rho_{xx}$ was 28 μΩ cm due to specific details of the thin film growth process. The giant $\xi_{DL}$ of 0.30 for the 4 nm Au$_{0.25}$Pt$_{0.75}$ is comparable to the high value reported for $\beta$-W [9] and 3 times higher than that of $\beta$-Ta [1]. One of the important factors that determines the non-monotonic $x$ dependence is the variation in $\rho_{xx}$ of the alloys. As can be clearly seen in Fig. 2(e), $\rho_{xx}$ of 4 nm Au$_{1-x}$Pt$_x$ thin film varies significantly and peaks at $x = 0.75$ due to enhanced electron scattering at that alloy concentration. We note that the maximum $\rho_{xx}$ is 83 μΩ cm for Au$_{0.25}$Pt$_{0.75}$, approximately one fourth that of $\beta$-W [9], which represents a considerable improvement of energy efficiency for SOT applications. $\xi_{FL}$ is negative and small in magnitude, relative to $\xi_{DL}$, and scales with $\xi_{DL}$, consistent with the SHE in Au$_{1-x}$Pt$_x$ alloys being the dominant source of both spin torques.

In Fig. 2(f) we plot the apparent spin Hall conductivity $\sigma^*_{SH} \equiv T_{int}\sigma_{SH} = (2e/\hbar)\mu_0 M_s t H_{DL} \rho_{xx}/j_e$ as a function of $x$ using the experimental results shown in Figs. 2(c) and 2(e). Assuming that $T_{int}$ is approximately independent of $x$ in the high Pt concentration regime, this plot indicates that $\sigma_{SH}$ remains constant or even initially increases slightly for decreasing $x$ in the Pt rich regime, $0.6 < x \leq 1.0$. Moreover, as also shown in Fig. 2(f), the measured $\sigma^*_{SH}$ for the 4 nm samples has a functional $x$ dependence similar to a recent calculation of $\sigma^{intr}_{SH}$ for bulk Au$_{1-x}$Pt$_x$ [13], which supports the conclusion that in this system the intrinsic spin Hall contribution arising from the topology of the electronic band structure is dominant.

### C. Spin diffusion length and spin Hall conductivity

The spin diffusion length is a key parameter for the theoretical understanding of a spin Hall material, for determining $\theta_{SH}$ via inverse SHE experiments and for optimizing SOT effectiveness. We determined $\lambda_s$ for Au$_{0.25}$Pt$_{0.75}$ by out-of-plane harmonic measurements of $H_{DL}$ and $\xi_{DL}$ for a series of Au$_{0.25}$Pt$_{0.75}$ $d$/Co 0.8 bilayers as a function of thickness $d$. The results plotted in Figs. 3(a) and 3(b) show that $H_{DL}$ and $\xi_{DL}$ increase as $d$ increases, as expected from the "bulk" SHE, with $\xi_{DL}$ saturating at $\approx 0.35$ by $d = 8$ nm. In Fig. 3(c) we plot the resistivity of the Au$_{0.25}$Pt$_{0.75}$ layers as a function of $d$ showing that $\rho_{xx}$ reaches its bulk value of 83 μΩ cm by $d \approx 3$ nm, consistent with a high resistivity material having a short bulk mean-free-path. In Fig. 3(d) we plot $\sigma^*_{SH}$ of the Au$_{0.25}$Pt$_{0.75}$ alloy. If we assume that the SOT arises entirely from the SHE of Au$_{0.25}$Pt$_{0.75}$, as indicated by the $d$ dependence of $\xi_{DL}$ and that the interfacial spin mixing conductance $G^{\uparrow\downarrow} \approx \text{Re}G^{\uparrow\downarrow} \gg \text{Im}G^{\uparrow\downarrow}$, we can obtain $\lambda_s$ and $\sigma_{SH}$ following $\sigma^*_{SH} = \sigma_{SH}(1-\text{sech}(d/\lambda_s))(1+\tanh(d/\lambda_s)/2\lambda_s \rho_{xx}\text{Re}G^{\uparrow\downarrow})^{-1}$ [17]. This assumes that $T_{int}$ is set by the spin backflow of an ideal Au$_{0.25}$Pt$_{0.75}$/Co interface with no interfacial spin-flip scattering (spin memory loss). We use $\rho_{xx} = 83$ μΩ cm and, as an approximation, $G^{\uparrow\downarrow} \approx 5.9 \times 10^{14}$ $\Omega^{-1}$m$^{-2}$, as calculated for the Pt/Co interface [17] to fit the data as shown by the solid line in Fig. 3(d). This fit ignores the slight variation in $\rho_{xx}$ for small $d$ and any change in $\lambda_s$ with that variation. From this fit we determine $\lambda_s \approx 1.7 \pm 0.1$ nm, $\sigma_{SH} \approx (7.0 \pm 0.1) \times 10^5$ $\Omega^{-1}$ m$^{-1}$, and an internal spin Hall ratio $\theta_{SH} = 0.58 \pm 0.01$. This value for $\lambda_s$ together with the measured $\rho_{xx}$ corresponds to a Au$_{0.25}$Pt$_{0.75}$ spin conductance $1/\lambda_s \rho_{xx} \approx 0.71 \times 10^{15}$ $\Omega^{-1}$m$^{-2}$, ~ 40% less than $1.25 \times 10^{15}$ $\Omega^{-1}$m$^{-2}$ obtained for pure Pt from a similar thickness dependent study [12]. Given the large value for $\sigma_{SH}$ obtained from this thickness-dependent measurement which as it assumes an ideal Au$_{1-x}$Pt$_x$/Co interface is perhaps just a lower bound, the proximity of the measured $\sigma^*_{SH}$ value for $d \approx 8$ nm to the theoretically calculated value [13] also shown in Fig. 2(d) is fortuitous. Instead our measurements are indicating that internal $\sigma_{SH}$ for Au$_{0.25}$Pt$_{0.75}$ is much larger than yet calculated, in accord with the results previously found for pure Pt [12]. Finally we note that the value for $\lambda_s$ here is considerably larger than ~0.2 nm (less than one atomic layer) used in the extraction of $\theta_{SH}$ in Au$_{1-x}$Pt$_x$/Ni$_{81}$Fe$_{19}$ bilayers from spin pumping and inverse SHE measurements [13]. This latter technique also does require a quite accurate determination of $G^{\uparrow\downarrow}$, which can be challenging to achieve, particularly if there is significant spin memory loss at the interface [16,18].

### D. Energy efficiency for spin-torque applications

Finally, we point out that Au$_{0.25}$Pt$_{0.75}$ is a particularly notable spin Hall material for SOT research and technological applications. As an example, we first show in Fig. 4(a) the deterministic switching of a 0.8 nm thick perpendicularly magnetized Co layer with a large PMA ($H_k \approx 6620$ Oe) and a high coercivity ($H_c \approx 800$ Oe) enabled by the giant $\xi_{DL}$ generated by the SHE of Au$_{0.25}$Pt$_{0.75}$. A small magnetic field of ± 50 Oe was applied along the current direction to overcome the Dzyaloshinshii-Moriya interaction at the Au$_{0.25}$Pt$_{0.75}$/Co interface. For technological applications, e.g. MRAMs, very efficient SOT switching of a FeCoB thin layer is of great interest. Figure 4(b) compares the write energy efficiency for SOT-MRAM applications based on various strong spin Hall channel materials whose $\xi_{DL}$, $\rho_{xx}$, and $\sigma^*_{SH}$ are listed in Table I. Here we calculate a MRAM device with a 400×200×4 nm$^3$ spin Hall channel and a 110×30×1.8 nm$^3$ FeCoB free layer (resistivity $\rho_{FeCoB} = 130$ μΩ cm) by taking into account the current shunting into the free layer (see the inset of Fig. 4(b)). The write power consumption for Au$_{0.25}$Pt$_{0.75}$ is 5, 8, 10 and 30 times smaller than that for Pt, Pt$_{0.85}$Hf$_{0.15}$, $\beta$-W, and $\beta$-Ta, respectively. The SOT-MRAMs based on topological insulators Bi$_2$Se$_3$ and (Bi,Se)$_2$Te$_3$ which are reported to have $\xi_{DL}$ of 0.16-3.5 [19,20] also have much higher power consumption than Au$_{0.25}$Pt$_{0.75}$ due to the considerable current shunting effect that is a consequence of their giant resistivity (e.g. 4020 μΩ cm for (Bi,Se)$_2$Te$_3$). We also note that Au$_{0.25}$Pt$_{0.75}$ is compatible with sputtering techniques and the use of Si substrates which are preferable for integration technology, whereas Bi$_2$Se$_3$ and (Bi,Se)$_2$Te$_3$ require costly molecular beam epitaxy growth and GaAs or sapphire substrates [19,20]. The high $\rho_{xx}$ of $\beta$-W, $\beta$-Ta, Bi$_2$Se$_3$, or (Bi,Se)$_2$Te$_3$ is also problematic for applications that require energy efficiency, e.g. the prospective implementation of SOT devices in cryogenic computing systems [11]. For example, use of a spin Hall material with a large resistivity will result in a high write impedance that is difficult for superconducting circuits in a cryogenic computing system to accommodate. Therefore, the combination of the giant $\xi_{DL}$, the relatively low resistivity, and the compatibility with sputtering and silicon integration technology makes Au$_{0.25}$Pt$_{0.75}$ a milestone spin Hall material for SOT research and technological applications.

### III. CONCLUSIONS

We have found Au$_{0.25}$Pt$_{0.75}$ to be a strong spin Hall material with a giant $\theta_{SH}$ of > 0.58 ($\xi_{DL}$ of 0.35 in bilayers with Co), a relatively low resistivity of ~83 μΩ cm, a large $\sigma_{SH}$ of > 7.0×10$^5$ Ω$^{-1}$ m$^{-1}$, and spin diffusion length of ~1.7 nm. This giant $\theta_{SH}$ and $\xi_{DL}$ arising from the intrinsic contribution of the SHE in Au$_{1-x}$Pt$_x$ alloys and the low resistivity make Au$_{0.25}$Pt$_{0.75}$ more energy-efficient than all the other conventional heavy metals and the topological insulators Bi$_2$Se$_3$ and (Bi,Se)$_2$Te$_3$ whose spin Hall properties have yet been reported. Our direct demonstration of the spin torque efficiency of this milestone spin Hall material Au$_{0.25}$Pt$_{0.75}$ which simultaneously combines a giant charge-spin conversion efficiency, low resistivity, and chemical stability, with excellent processing compatibility for device integration, paves the way for further fundamental research that will benefit from the efficient generation of spin currents and of spin-orbit torques, and for the development of new highly efficient SOT-driven magnetic memories, skyrmion and chiral domain wall devices, and microwave and terahertz emitters.

We thank S. Shi, Y. Ou, and R. C. Tapping for discussion. This work was supported in part by the Office of Naval Research, by the NSF MRSEC program (DMR-1719875) through the Cornell Center for Materials Research, and by the Office of the Director of National Intelligence (ODNI), Intelligence Advanced Research Projects Activity (IARPA), via contract W911NF-14-C0089. The views and conclusions contained herein are those of the authors and should not be interpreted as necessarily representing the official policies or endorsements, either expressed or implied, of the ODNI, IARPA, or the U.S. Government. The U.S. Government is authorized to reproduce and distribute reprints for Governmental purposes notwithstanding any copyright annotation thereon. This work was performed in part at the Cornell NanoScale Facility, an NNCI member supported by NSF Grant ECCS-1542081.

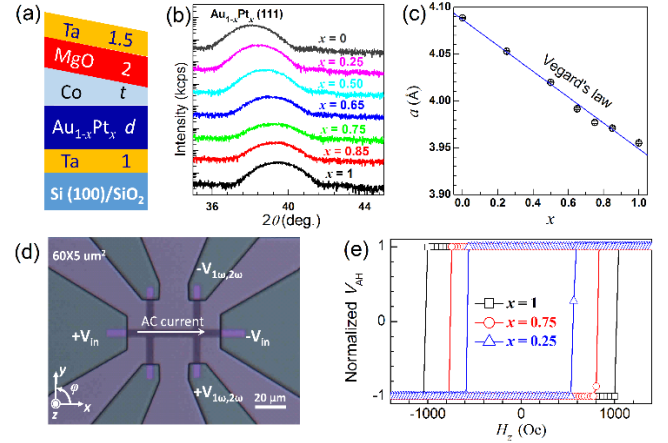

FIG. 1. (a) Schematic depiction of the magnetic stacks with $t$ and $d$ being the thicknesses of Co and $Au_{1-x}Pt_x$ layers (the thicknesses are in nm); (b) X-ray diffraction patterns for different $Au_{1-x}Pt_x$ composition $x$; (c) Optical microscopy image of a Hall bar device showing the geometry and the measurement coordinate; (d) Normalized anomalous Hall voltage for $Au_{1-x}Pt_x$ 4/Co 0.8 bilayers ($x = 1$, 0.75, and 0.5, respectively).

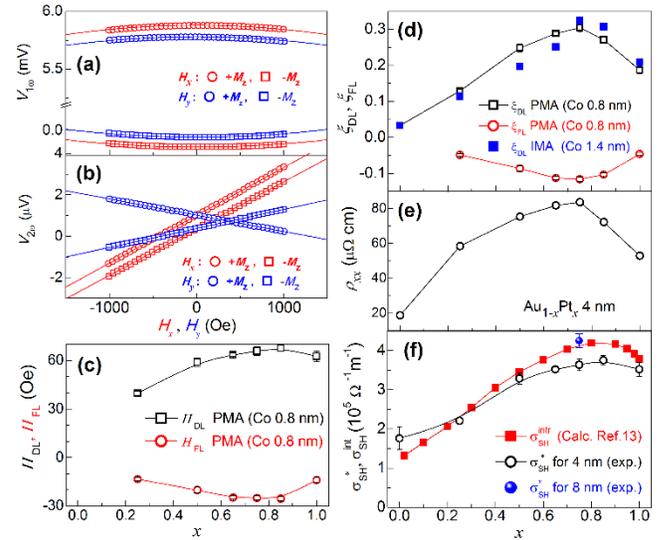

FIG. 2. Spin-orbit torques in $Au_{1-x}Pt_x$/Co bilayers. (a) The first ($V_{1\omega}$) and (b) second ($V_{2\omega}$) harmonic voltages plotted as a function of in-plane fields $H_x$ (red) and $H_y$ (blue), respectively (for $x = 0.75$); Composition dependence of (c) effective spin-torque fields $H_{DL}$ and $H_{FL}$, (d) spin-torque efficiencies $\xi_{DL}$ and $\xi_{FL}$, (e) $Au_{1-x}Pt_x$ resistivity $\rho_{xx}$, and (f) *calculated* intrinsic spin Hall conductivity $\sigma_{SH}^{intr}$ for bulk $Au_{1-x}Pt_x$ (red squares) and experimentally measured effective spin Hall conductivity $\sigma_{SH}^* \equiv T_{int}\sigma_{SH}$, where $T_{int} \cong 0.6$ if there is no spin memory loss at the $Au_{1-x}Pt_x$/Co interfaces, for 4 nm (black circle) and 8 nm $Au_{0.25}Pt_{0.75}$ (blue solid dot), respectively. For clarity, the $H_y$-dependent $V_{1\omega}$ data (the blue points) in (a) are artificially shifted by ±0.001 mV for $\mp M_z$, respectively.

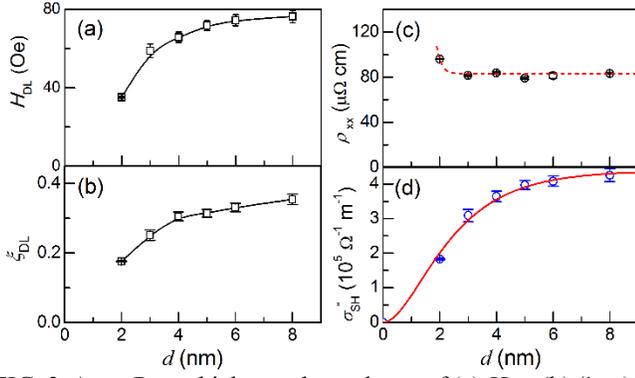

FIG. 3. $Au_{0.25}Pt_{0.75}$ thickness dependence of (a) $H_{DL}$, (b) $\xi_{DL}$, (c) $\rho_{xx}$, and (d) $\sigma_{SH}^*$, respectively.

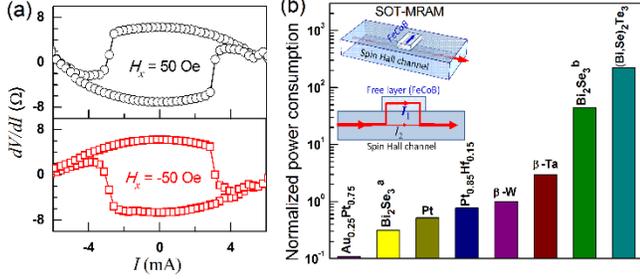

FIG. 4. Spin-orbit torque switching of magnetetization. (a) Deterministic direct current switching of a 0.8 nm perpendicualrly magnetized Co layer with coercivity of 800 Oe and perpendicular anisotropy field of 6620 Oe via the damping-like SOT geneareted by the giant SHE of a 4 nm $Au_{0.25}Pt_{0.75}$ layer. Magnetic fields of ± 50Oe were applied along the current direction, respectively. (b) Normalized power consumption as calculated for a typical SOT-MRAM device with different spin Hall channel materials listed in Table 1. Insets: schematics of a SOT-MRAM device and its side-view highlighting the increase of the write current due to the current shunting into the magnetic free layer.

TABLE I. Comparison of $\xi_{DL}$, $\rho_{xx}$, and $\sigma_{SH}^*$ of various spin Hall materials.

|  | $\xi_{DL}$ | $\rho_{xx}$ ($\mu\Omega$ cm) | $\sigma_{SH}^*$ ($10^5$ $\Omega^{-1}$ m$^{-1}$) | Ref. |
|---|---|---|---|---|
| $\beta$-Ta | 0.12 | 190 | 0.63 | [1] |
| $\beta$-W | 0.3 | 300 | 1.0 | [9] |
| Pt | 0.12 | 50 | 2.4 | [12] |
| $Pt_{0.85}Hf_{0.15}$ | 0.16 | 110 | 1.5 | [10] |
| $Bi_2Se_3$[a] | 3.5 | 1755 | 2.0 | [18] |
| $Bi_2Se_3$[b] | 0.16 | 1060 | 0.15 | [19] |
| $(Bi,Se)_2Te_3$ | 0.4 | 4020 | 0.1 | [19] |
| $Au_{0.25}Pt_{0.75}$ | 0.35 | 80 | 4.4 | This work |